\documentclass[12pt]{iopart}
\usepackage{graphicx}
\begin{document}

\title{Dynamical and spectral properties of complex networks}


\author{Juan A. Almendral$^1$ and Albert D\'{\i}az-Guilera$^{2,3}$}
\address{$^1$ Departamento de F\'{\i}sica, Universidad Rey Juan Carlos, 28933 M\'ostoles, Madrid, Spain}
\address{$^2$ Departament de F\'{\i}sica Fonamental, Universitat de Barcelona, 08028 Barcelona, Spain}
\address{$^3$ Institute for Biocomputation and Physics of Complex Systems (BIFI), University of Zaragoza, Zaragoza 50009, Spain}
\ead{juan.almendral@urjc.es}
\ead{albert.diaz@ub.edu}

\begin{abstract}
Dynamical properties of complex networks are related to the spectral properties of the Laplacian matrix that describes
the pattern of connectivity of the network. In particular we compute the synchronization time for different types of
networks and different dynamics. We show that the main dependence of the synchronization time is on the smallest
nonzero eigenvalue of the Laplacian matrix, in contrast to other proposals in terms of the spectrum of the adjacency
matrix. Then, this topological property becomes the most relevant for the dynamics.
\end{abstract}


\submitto{\NJP}

\section{Introduction}

In the last decade we have witnessed an enormous effort in understanding the complex patterns of connectivity that has
been found in many natural, social or technological systems~\cite{s01,ab02,dm02,n03,boccaletti06}. Once the systems are
characterized from a topological point of view, it is the turn of the dynamical properties and relating both dynamic
and static characterizations has become one of the hot topics in network theory in the last years. There can be many
different dynamics implemented in networks, ranging from simple discrete state systems, like cellular automata or
random boolean networks, to networks of units whose individual behaviour is already complex, as it happens in samples
of coupled chaotic units. But, when dealing with the emergent behaviour characteristic of complexity, one of the main
issues is to discern between the effects related to the topology and the effects related to the dynamical rules of the
units.

In this paper we want precisely to understand which is the main topological characteristic of a network (undirected and
unweighted) that influences the dynamical response. By looking at particular dynamical rules of the individual nodes
and at particular rules of interaction between the units, the goal is to see what do they have in common in terms of
the static properties of the network. In particular we want to analyze the route of complex networks to
synchronization, understood as a stationary state in which all the units are in the same state. Synchronization of
complex networks has been widely analyzed during the past years~\cite{mmko06}, mainly in the context of the Master
Stability Formalism (see~\cite{bp02,dhm05,mzk05a,mzk05b,zmk06,nm06a,dnm06}) that studies the stability of the
synchronized state. Other studies have focused on the behaviour of small structures (motifs), as for
instance~\cite{mvp04,lbl06}, and some other recent publications analyze the behaviour along the complete evolution of
the system~\cite{adp06a,adp06b,ad06,gma06,lsas06,motter07}, highlighting the relation between spectral, topological and
dynamical properties of networks.

Synchronization is a general concept and not related to a particular type of dynamics; for this reason, in order to
stress on the dynamical significance of the network parameters we consider three different types of dynamical evolution
of the units and of the interaction rules: linear dynamics as a general approximation when the variables describing the
units state are close to each other and hence close to the synchronized state, Kuramoto dynamics~\cite{abprs05} that
have been widely used in physical and biological problems, and a generic model of spin-like units that could model
interactions between individuals in a social network.

The outline of the paper is as follows. In Sect.~2 we analyze the simplest case of linear interactions between units.
In Sect.~3 we compare the roles of the eigenvalues of the adjacency matrix and the Laplacian matrix, going back to the
dynamics in Sect.~4 where the two different nonlinear dynamical rules are considered. We end in Sect.~5 with the
conclusions of our work.

\section{Linear dynamics}

Synchronization is a generic feature of coupled identical dynamical systems. In the last years the physicists community
has been focusing in the effect of connectivity patterns that go far beyond the usual regular lattices or all-to-all
neighboring schemes. Nowadays we have become used to consider units as nodes of a graph that are linked to other units
in a complex way. Each unit evolves following its own dynamics and they are coupled according to some rules. Under
generic conditions the coupling tends to favor the synchronization of the units. The first theoretical attempt to
analyze the stability of the synchronized state of a complex network was made by Barahona and Pecora~\cite{bp02}.
Keeping the formalism to a minimum they proposed a system that obeys the following set of equations of motion

\begin{equation}
\frac{dx_i}{dt}=F(x_i)-\sigma\sum_{j=1}^{n}L_{ij}H(x_j)
\end{equation}
where $F$ corresponds to the unit evolution and $H$ stands for the coupling; $\sigma$ is the coupling strength and
$L_{ij}$ is the Laplacian matrix, related to the adjacency matrix, $A_{ij}$ by the following relation:
\begin{equation}
L_{ij}= k_i \delta_{ij} - A_{ij}. \label{laplacian}
\end{equation}
This Laplacian matrix is symmetric with zero row-sum and hence all the eigenvalues are real and non-negative. The
eigenvalues are ordered such that
\begin{equation}
0=\lambda_1\leq\lambda_2\leq...\leq\lambda_N.
\end{equation}
The number of zero eigenvalues is equal to the number of connected components. If we are concerned with synchronization
as a global effect, we have to notice that it is only possible in systems with a single connected component and hence
there will be a single zero eigenvalue, implying $\lambda_2>0$. In general, the following inequality is also
fulfilled~\cite{am85}
\begin{equation}
\lambda_N\leq 2k_{max}
\end{equation}
where $k_{max}$ is the largest degree in the graph.

Barahona and Pecora show that the synchronized state is stable if $\frac{\lambda_N}{\lambda_2} <
\frac{\alpha_B}{\alpha_A}$, where ${\alpha_A}$ and ${\alpha_B}$ are the lower and upper bounds, respectively, of the
effective coupling $\sigma\lambda_i$ in which the maximum Liapunov exponent is negative. This inequality involves a
part that depends on the topology of the network, the eigenvalue ratio $Q=\frac{\lambda_N}{\lambda_2}$, and a part that
depends on the dynamical properties of the functions $H$, $F$, and the values of the variables in the synchronized
state. Thus it could be concluded~\cite{dnm06} that the synchronizability of the system, understood as the stability of
the synchronized state, is enhanced if the ratio $Q$ is as small as possible. Since the value of the largest eigenvalue
$\lambda_N$ depends mainly on the maximum degree of the network, the main dependence will be, according
to~\cite{dnm06}, on the smallest nonzero eigenvalue $\lambda_2$, usually called the \emph{spectral gap}.

In this paper we want to perform an additional step in the direction of characterizing synchronization in complex
topologies, and for this reason we propose, as an additional parameter, the time the system needs to synchronize.
Obviously, this characterization will depend on many factors: the type of dynamics of the single units, how strong is
the coupling, and how far is the initial setup from the synchronized state. In order to simplify this picture and to
analyze how this time depends on the topological properties we will consider the simplest case of dynamics and
coupling, and in the next sections we will study more complex dynamics. We will assume that each unit has a constant
driving $F(x_i)=\mbox{constant}$, the same for all the units, and hence we can fix it to zero by transforming to a
moving frame of reference. Furthermore, we will consider that the coupling is linear, which is a good approximation
when the values of the variables describing the system are close to each other. Thus we deal with the system of
differential equations
\begin{equation}
\frac{d\theta_i}{dt}=-\sigma\sum_{j} L_{ij} \theta_j \hspace{0.5cm} i=1,...,N. \label{linearmodel}
\end{equation}

We should remark that, although we introduce this set of differential equations as a first approach to the problem of
synchronization in complex networks, (\ref{linearmodel}) itself is interesting in the context of distributed systems
where it is known as \emph{consensus dynamics}~\cite{lynch97}, having a long history in the field of Computer Science.

The Laplacian matrix is related to the topological properties of the network and hence it is the only relevant
dependence; since the coupling strength $\sigma$ just fixes the time scale. There is another obvious dependence on the
initial conditions that will be discussed later.

The solution of this system reads in terms of the normal modes $\varphi_i$~\cite{adp06a}
\begin{equation}
\sum_{j} B_{ij}\theta_j=\varphi_i(t)= \varphi_i(0) e^{-\lambda_i t}\hspace{0.5cm} i=1,...,N \label{linearsolution}
\end{equation}
where $B_{ij}$ is the matrix of the transformation from the old coordinates to the new ones. Thus we are left with
linear combinations of phases in the original coordinates that is equal to a term that depends on the initial
conditions multiplied by an exponential that decays very fast in time according to the eigenvalues of the Laplacian
matrix. For very large $t$ the exponentials decay to zero and the only solution is that all the units become
synchronized. We can then assume that, at large times, the phase difference in the original coordinates decays
exponentially with the smallest eigenvalue $\lambda_2$.

Thus we can write
\begin{equation}
\theta_i - \theta_j = C e^{-\lambda_2 t}
\end{equation}
where $C$ is an unknown constant that depends on the specific details of the network and on the initial configuration.

Formally, the time the system needs to achieve complete synchronization is infinite. Usually in computer simulations
one establishes a relaxed synchronization condition. We say that two oscillators are synchronized if the cosine of
their phase difference is very close to 1,
\begin{equation}
\cos(\theta_i - \theta_j)\ge 1-\varepsilon, \label{threshold}
\end{equation}
which means that $\theta_i - \theta_j \sim \varepsilon^{1/2}$ and we can write
\begin{equation}
\frac{1}{2} \ln \varepsilon = \ln C -t \cdot \lambda_2
\end{equation}
and from here we can say that the synchronization time behaves in the following way:
\begin{equation}
T_{\mbox{\footnotesize{sync}}} \sim \frac{1}{\lambda_2} \left[ \ln C-\frac{1}{2}\ln\varepsilon \right]. \label{t_sync}
\end{equation}
It is clear that this time depends on the topology and on the threshold condition.

In order to check these statements we have performed numerical simulations of~(\ref{linearmodel}) for different
networks and thresholds. In all cases we have assumed random initial conditions in the range $[0,2\pi]$. \footnote{This
choice is due to the fact that in Section 4 we will deal with phase oscillators where this is the natural choice.} It
is precisely this dependence on the initial conditions that makes the synchronization time
$T_{\mbox{\footnotesize{sync}}}$ a fluctuating magnitude. The range in the initial conditions is the responsible of the
dispersion in the synchronization time, and this dispersion cannot be reduced by increasing the number of realizations.

\begin{figure}
\centering
\includegraphics*[width=0.7\textwidth]{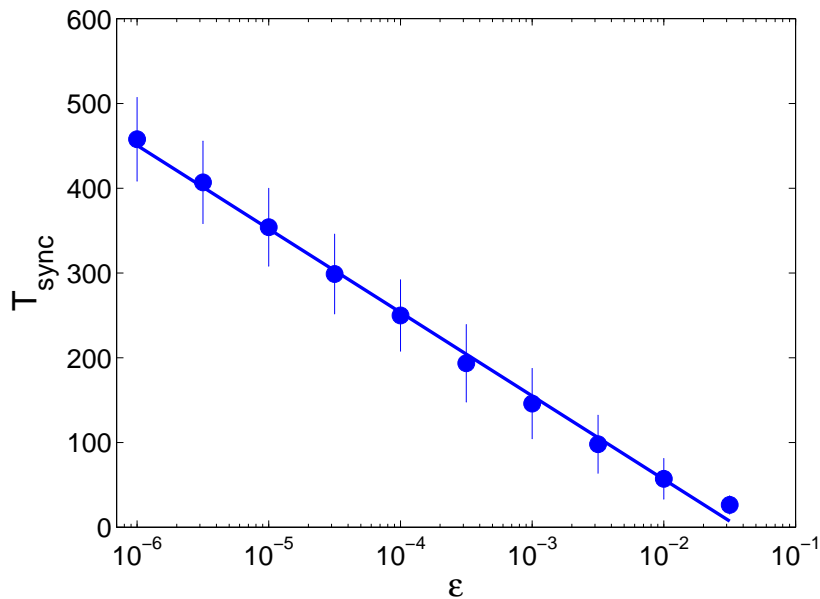}

\caption{Time to synchronize as a function of $\ln \varepsilon$. We have used a network of 256 nodes that was proposed
in~\cite{adp06a} as an example of network with two hierarchical community levels.} \label{fig_epsilon}
\end{figure}

We have considered different types of networks, with a wide range of topological features and sizes, just to focus on
the dependence on the relevant characterization of the dynamical response of the network. Before entering on the
details of the topology let us focus on the threshold dependence. To this purpose we consider a particular network and
change the synchronization condition~(\ref{threshold}). The results of this set of simulations is plotted in
Fig.~\ref{fig_epsilon}, where we can observe a clear linear dependence of the synchronization time on $\ln
\varepsilon$, thus providing support to the assumptions we have made before. We have checked this dependence on other
networks and dynamics and the conclusions  are the same, the main dependence on the threshold is of the type shown
in~(\ref{t_sync}).

Following these assumptions now we want to check that, fixing the threshold condition, the synchronization time depends
on the inverse of the spectral gap only. For this reason we have used networks with different sizes (128, 256 and 512),
community structure at different hierarchical levels (1 level, 2 levels, and no communities), and growing rules. We
have chosen networks grown according to different rules. Erd\"{o}s-R\'{e}ny (ER) random graphs~\cite{er59}; small-world
models as proposed by Newman and Watts (NW)~\cite{nw99}, in which the shortcuts are added instead of rewired as in the
original Watts-Strogatz model~\cite{ws98}; and Barab\'{a}si-Albert (BA) preferential attachment growing model
networks~\cite{ba99}.

\begin{figure}
\centering \includegraphics*[width=0.7\textwidth]{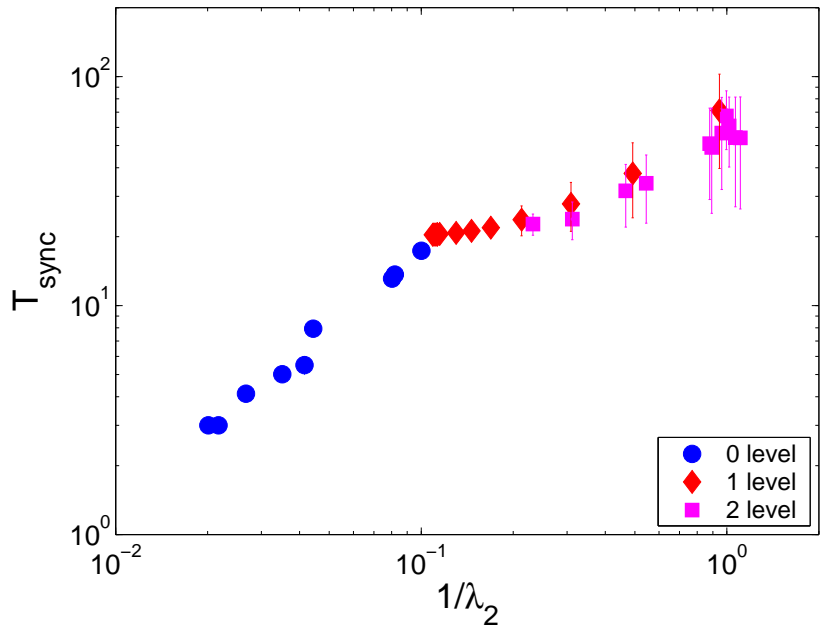}

\caption{Synchronization time for different networks. To show the dependence on the smallest non-zero eigenvalue of the
Laplacian matrix, we have plotted the time to synchronize as a function of $1/\lambda_2$. In these plots we fix the
threshold to $0.99$. We have used three sets of networks: networks introduced in~\cite{adp06a} that have two community
levels; networks introduced in~\cite{ng04} that have one community level; and networks with no community structure
grown with different rules (see main text for details).} \label{fig_time_lambda2}
\end{figure}

Although these networks have different features, and this is reflected in its synchronization time, we want to stress
that there is a clear dependence on the spectral gap, and this is indeed what can be concluded from
Fig.~\ref{fig_time_lambda2}, in which all the networks show an almost linear dependence on $1/\lambda_2$. Furthermore,
although when moving from one class of network to another class of network this dependence is not so clear, what one
can undoubtedly conclude is that there is a monotonic increase of the synchronization time on the spectral gap.
Networks with community structure need special care because neatly defined communities are related to very precise time
scales for the internal synchronization within the community (see~\cite{adp06a}) and, consequently, they show a
different slope in Fig.~\ref{fig_time_lambda2}.

This conclusion about the monotonic dependence on the spectral gap supports the previous analysis in~\cite{dnm06} that
highlights the role of this particular eigenvalue in the dynamical characterization of a complex network. In particular
these authors show, apart from the stability of the synchronized state, that random walks propagate more easily in
networks with large spectral gaps. This observation enables the authors to construct optimal graphs where the
optimization goal is precisely the lowest spectral gap. In any case, we have shown that synchronization time depends
mainly on this value for a set of linearly coupled dynamical systems. In Sect. 4 we will come back to this issue
dealing with other non-linear dynamics but, previously, we will discuss in next section the role played by this
eigenvalue and other proposals in the literature of complex networks.

\section{Spectral analysis}

In the previous section we have mentioned that, in recent literature, complex networks have been dynamically
characterized by the spectral gap in terms of the stability of the synchronized state, following the original arguments
of Barahona and Pecora~\cite{bp02}, and also in terms of random walks~\cite{dnm06}. We have provided additional support
by showing how the synchronization time strongly depends on this property. But also in recent literature some focus has
been put on slightly different characterizations, in particular in~\cite{roh06} the authors propose the largest
eigenvalue of the adjacency matrix. We have computed this eigenvalue $\lambda_N^A$ and compared it with the spectral
gap (the first non-zero eigenvalue of the Laplacian matrix). In general, for a complex network, there is no simple
relation between the eigenvalues of the two matrices, the adjacency matrix and the Laplacian matrix. Only in the
particular case of a regular lattice, or a network in which all nodes have exactly the same degree ($k$), the
eigenvalues satisfy the following relationship
\begin{equation}
\lambda_i^L=k-\lambda_{N-i+1}^A
\end{equation}
as can be easily concluded from~\ref{laplacian} and keeping the same ordering for the eigenvalues of the adjacency
matrix
\begin{equation}
\lambda_1^A\leq\lambda_2^A\leq...\leq\lambda_N^A.
\end{equation}
But if the distribution of degrees is not homogeneous, as it usually happens in complex networks, then the relation is
unknown. As a first approximation, for distributions of connectivities that are not far from the homogeneous one, as it
happens for instance in random and in small-world graphs, we can still consider it. In this case, since the first
eigenvalue of the Laplacian matrix is zero the largest eigenvalue of the adjacency matrix should be close to the
average degree. Thus, networks that are quite homogeneous in degree will have a value of $\lambda_N^A$ that is very
close to the average degree, and hence it provides little information about the network structure and its dynamical
properties. For this reason our proposal of characterizing the dynamical response of the system by the spectral gap is
more appropriate.

\begin{figure}
\centering \includegraphics*[width=0.7\textwidth]{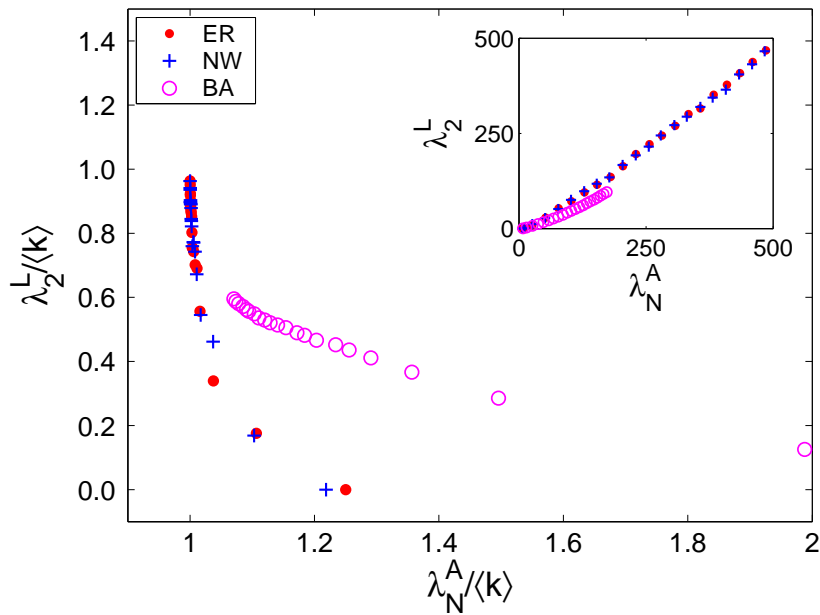}

\caption{Smallest nonzero eigenvalue of the Laplacian matrix versus the largest eigenvalue of the adjacency matrix for
different networks, both of them divided by the average degree of each network. Inset: the same but without
normalization. We have chosen networks grown according to different rules. ER stands for Erd\"{o}s-R\'{e}ny random
graphs \cite{er59}; NW for Newman-Watts~\cite{nw99}, in which the shortcuts are added instead of rewired as in the
original Watts-Strogatz model~\cite{ws98}; and BA for Barab\'{a}si-Albert preferential attachment growing
model~\cite{ba99}.} \label{lambdas}
\end{figure}

In Fig.~\ref{lambdas} we plot the second eigenvalue of the Laplacian matrix ($\lambda_2^L$) against the largest
eigenvalue of the adjacency matrix ($\lambda_N^A$). There we can observe several facts that deserve some comments.
First, for networks with a homogeneous distribution of degrees, the largest eigenvalue of the adjacency matrix divided
by the mean degree of the graph shows a very slight dependence on the network structure. On the contrary,  the first
nonzero eigenvalue of the Laplacian, divided by the mean degree as well, presents a more pronounced dependence on the
network structure. Second, for networks with an inhomogeneous distribution of degree, such as the ones grown with the
Barab\'{a}si-Albert preferential attachment rule, both eigenvalues change with the network under consideration but they
change in a similar fashion conserving a linear relationship. Then, one can conclude that, even in this case, there is
no additional information in the adjacency matrix with respect to the Laplacian one. In summary, the first nonzero
eigenvalue of the Laplacian matrix is more sensitive to network changes than the largest eigenvalue of the adjacency
matrix, and for this reason it will be the focus of the next sections.

\section{Nonlinear dynamics}

In order to check that the conclusions we have drawn before about the characterization of the dynamical properties of
complex networks in terms of the spectral gap goes beyond the linear model, in this section we are going to consider
two quite different nonlinear models: a model of phase oscillators and a model of spins. In both cases, according to
the dynamics and the interaction, the system tends to synchronize. One of them synchronizes in a threshold sense, since
phase is continuous, and the other in an absolute way since the spin states are discrete.

\subsection{Kuramoto oscillators}

One of the most studied model of coupled oscillators is that due to Kuramoto~\cite{abprs05}. In this case oscillators
follow the dynamics:
\begin{equation}
\frac{d\theta_i}{dt}=\omega_i + \sigma\sum_{j} A_{ij}\sin(\theta_j-\theta_i) \hspace{0.5cm} i=1,...,N \label{ks}
\end{equation}
\noindent where $\omega_i$ stands for the natural frequency of the oscillator and $\sigma$ describes the coupling
between adjacent units. If the oscillators are identical $(\omega_i = \omega ~\forall i)$ there is again only one
attractor of the dynamics: the fully synchronized regime where $\theta_i = \theta, ~\forall i$, which is stable. There
has been a lot of effort in the recent literature on this model applied to complex networks,
\cite{mp04,adp06a,adp06b,ad06}. Concerning our current discussion relating spectral and dynamical properties one should
notice~\cite{adp06a} where the intermediate time scales are related to the topological scales of the networks which, in
turn, are related to the distribution of eigenvalues. As we said before, the number of zero eigenvalues of the
Laplacian matrix is equal to the number of connected components of the network. It is trivial then to conclude that if
$\lambda_2^L=0$, the network is split in more than one disconnected subnetworks. Then, from a dynamical point of view,
it is impossible for the network to achieve a complete synchronized state, being only possible subnetworks with
internal coherence but not synchronized between them. Thus, a small value means that we are close to this situation and
that it will take a long, although finite, time to synchronize completely (no matter how close to 1 the threshold
condition is chosen).

Actually, in~\cite{adp06a,adp06b} is shown that the existence of clearly defined communities~\cite{ddda05}, groups of
nodes in which the number of internal links is large compared with the number of external ones, is related to gaps in
the values of the eigenvalues, and the order of the gap is related to the number of communities, the sharper the
community definition the larger the gap. This relation between spectral properties of the Laplacian matrix and the
topological properties of the network is in turn reflected in the dynamics. Starting from a random distribution of
phases and averaging over a set of distributions, what was observed is that synchronization appears from the innermost
local scale to the outermost (albeit global) scale. In this synchronization process, the groups of nodes that get
synchronized correspond to the topological communities and the times at which the groups merge to form larger groups
are related to gaps in the spectrum of the Laplacian matrix. This relation closes the interdependence between
topological, spectral and dynamical properties of the network. Then, in this general framework in which gaps in the
spectrum are related to the achievement of synchronization at different scales, the last gap $\lambda_2$, always large
compared with $\lambda_1=0$, should correspond to the completion of the synchronization process at the largest global
scale.

\begin{figure}
\centering \includegraphics*[width=0.7\textwidth]{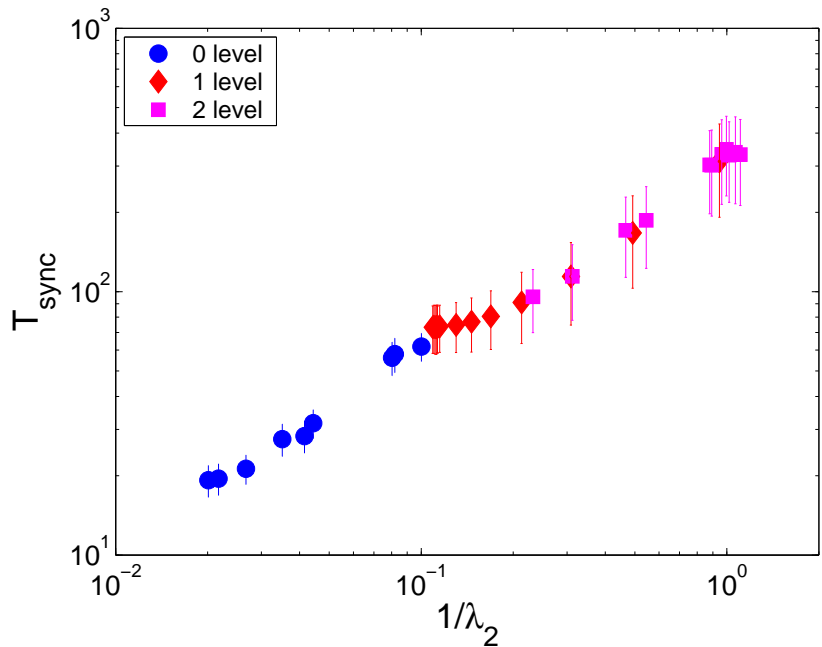}

\caption{Synchronization time for the same networks than in Fig.~2, for dynamical evolution given by~\ref{ks}. All the
details of the simulations, such as distribution of random phases, threshold for synchronization, and coupling
constant, are the same.} \label{kura_fig}
\end{figure}

With this goal in mind, we are going now to evaluate the synchronization time for a system of Kuramoto oscillators and
analyze its dependence on the spectral gap. Before entering into the simulation details one has to notice that,
starting the dynamics from a random distribution of phases, the oscillators rapidly settle into closer phases; after
this fast initial evolution all phases are quite similar and the sine function in~\ref{ks} can be well approximated by
its argument. For this reason the linear model discussed in Sect.~2 is a good approximation for the Kuramoto
oscillators at later times, close to the synchronized state. This does not ensure, however, that this happens along all
time evolution of the system and one has to be careful if the evolution concerns all time scales. In our numerical
simulations we have analyzed the same type of networks than in Sect.~2 for the Kuramoto dynamics and the results are
shown in Fig.~\ref{kura_fig}.

In Fig.~\ref{kura_fig} we can see that the time to synchronize for the different networks is around one order of
magnitude larger; this is of course due to the initial rearrangement of the oscillators. Whereas in the linear model
they are coupled linearly, in the Kuramoto model they are coupled through the sine function, which is smaller and makes
the transient time needed to get similar phases larger. Once this transient is over, the phases are very similar and
the two models behave in exactly the same way. Of course, if initial conditions in a small interval were chosen, the
difference with respect to the linear model would be shortened, and the transient time reduced.
In any case, the important notice is that the scaling of the synchronization time with the spectral gap is identical in
both models, enhancing our assumptions about the importance of the spectral gap as the key characterization of the
dynamical response of the system. As it happens with the linear interaction rules, here the dependence on the inverse
of the spectral gap is not linear for the whole set of networks.
Although the internal structure of the networks with communities can introduce important effects in the route towards
global synchronization and break slightly the linear dependence, the monotonously increasing behaviour with
$1/\lambda_2$ is maintained.

\subsection{Majority dynamics}

The linear model analyzed in Sect.~2 and the Kuramoto model discussed in the previous paragraph are described by
continuous variables, phases, and the synchronization is understood in a threshold sense. In any case we have shown
that the synchronization process is similar in both cases, being quite different in the transient to synchronization
but very similar when arriving to the synchronized state. In this subsection we propose a completely different model,
in which the dynamics is discrete, and hence not sharing the processes described above.

Let us consider a discrete spin-like system in which the nodes of a network have only two possible states, $s_i= \pm
1$. This could model, for instance, the dynamics of public opinion in social influence networks (e.g. when a group of
people choose among two different opinions). Initially, half of the spins are randomly set at the state $-1$ and the
other half at $+1$. Then, each node $i$ receives an input $h_i = \sum_j A_{ij} s_j$, being ${A}_{ij}$ the adjacency
matrix. In this manner, as other authors have pointed out~\cite{w99,zl05}, this spin-like network is characterized at
each step by some pattern of internal states, whose evolution represents the global dynamics.

We evolve the network according to the following local majority rule: the state of node $i$ at time $t+1$ is given by
\begin{equation}
s_i(t+1) = \left\{
\begin{array}{cc}
+1 & \mbox{if } h_i(n) > 0 \\
s_i(t) & \mbox{if } h_i(n) = 0 \\
-1 & \mbox{if } h_i(n) < 0
\end{array}
\right.
.
\end{equation}

We find that the system does not synchronize for some initial states. That is, the system wanders in the phase space
without reaching a fixed point. This phenomenon is a kind of frustration in which the system is unable to reach the
lowest energy state.

To focus our attention on how the topology contributes to synchronize the system, we overcome this frustration by
introducing a slight perturbation, which can be regarded as a noise or a thermal bath. We find that it is enough, for
the system to synchronize, that $0.5\%$ of the states are randomly switched at every step. The introduction of such a
perturbation has a drawback, in the sense that it also destroys the final synchronized state. For this reason we will
consider that the system is synchronized if $99\%$ of the nodes are in the same discrete state.

According to these rules we have performed several numerical simulations to compute the average time required to
synchronize networks, all of them formed by 512 nodes, grown again following different models. Namely, we have
considered several Erd\"{o}s--R\'{e}nyi (ER), Newman--Watts (NW) and Barab\'{a}si--Albert (BA) networks.

Since it is impossible to synchronize a system with disconnected components, we have only studied networks with a
unique connected component and refused those that do not verify such condition. This is not a problem for NW and BA
networks because their growth is such that all nodes are linked. But this is not the case for ER networks in general
and we have only considered ER networks with a single component.

\begin{figure}
\centering \includegraphics*[width=0.7\textwidth]{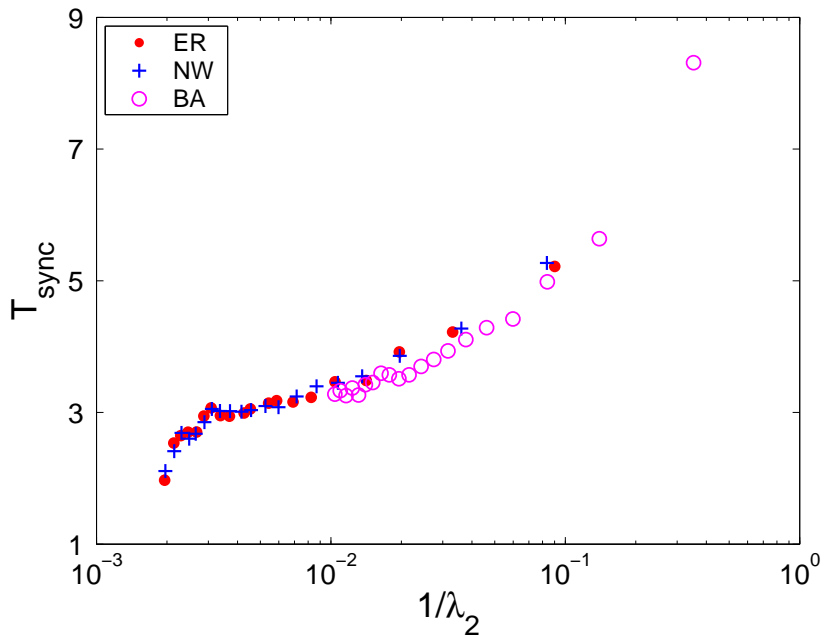}

\caption{Synchronization time for a discrete spin-like dynamics following a local majority rule. Three types of
networks are considered: Erd\"{o}s--R\'{e}nyi, Newman--Watts, and Barab\'{a}si--Albert. In all of them we find that the
synchronization time monotonically grows with $1/\lambda_2$.} \label{spin}
\end{figure}

We observe that the synchronization time monotonously depends on the inverse of the spectral gap as we have found in
the previous simulations (Fig.~\ref{spin}). It is then clear again the importance of this particular eigenvalue,
although in the present case the functional dependence is different from the previous  continuous models (linear and
Kuramoto). Nevertheless, due to the fact that the dynamical rules are completely different, it gives more arguments to
our line of reasoning and the spectral gap should be considered the main characteristic of the network concerning the
dynamical response.

In principle, the dynamical evolution of this spin-like system is quite different from the linear and Kuramoto models
analyzed above. However, the dynamical evolution can be written in compact form
\begin{equation}
s_i(t+1)=\Theta\left[ \sum_{j}A_{ij}s_j(t)+\mu s_i(t) \right]
\end{equation}
where $\Theta (x)$ is the Heaviside step function, and $\mu$ is a number $0<\mu <1$ that ensures that there is no
change in case of tie between the neighbours. In this equation one can notice that the argument of the Heaviside
function can be written in matrix form
\begin{equation}
\sum_{j}A_{ij}s_j(t)+\mu s_i(t) = \sum_{j} \left[ A_{ij}+\frac{\mu}{k_i}\delta_{ij}\right] s_j(t)
\end{equation}
In this way the evolution of the units reads
\begin{equation}
s_i(t+1)=\Theta\left[  \sum_{j} \left[ A_{ij}+\frac{\mu}{k_i}\delta_{ij}\right] s_j(t) \right]
\end{equation}
in which two important facts should be noticed. First, the Heaviside function imposes a quite strong nonlinearity that,
eventually, could be regularized but, in principle, can be the responsible for the rapid convergence towards the
synchronized state of this dynamical rule. Second, the matrix in the argument
\begin{equation}
A_{ij}+\frac{\mu}{k_i}\delta_{ij}
\end{equation}
plays a key role in the way the coupling is performed. This matrix, which can be related to the Laplacian and adjacency
matrices, can be analyzed in the same terms than those. Since its eigenvalues should give information about the
dynamical processes taking place, its calculation and comparison with the other spectral properties becomes relevant
and it will be the objective of a future work.

\section{Conclusions}

In this paper we have presented results on different types of dynamics running in complex networks. We propose a new
dynamical measure to characterize the dynamical properties of networks, the synchronization time. Although this time
can depend on many factors, mainly the type of dynamics that is implemented and on others like the coupling and the
initial conditions, we observe that this time basically depends on one of the static features of the network, the so
called spectral gap, the smallest nonzero eigenvalue of the Laplacian matrix. The role played by this eigenvalue has
also been stressed in the features of complex networks related to the dynamics. We have also compared this eigenvalue
with other proposal of the literature, based on the spectral properties of the adjacency matrix, and we have found that
the dependence is clearer in terms of the inverse of the spectral gap of the Laplacian matrix. Likewise, we have found
that different dynamics can be described in terms of other matrices, different from the traditional studies based on
the Laplacian or adjacency matrices.

The study we have performed relating the spectral, topological and dynamical properties of complex networks has an
immediate continuation in terms of the robustness of the network. Usually robustness is defined in terms of the
topology, i.e. how the network connectivity responds to external attacks, but we are convinced that relating
topological and dynamical properties would give more hints on the dynamical robustness of the network, which is the
dynamical response of the network to dynamical attacks.

\ack
Fruitful discussions with A. Arenas and I. Sendi\~{n}a-Nadal are appreciated. Financial support from projects
FIS2006-13321 and  FIS2006-08525 of the Spanish Government, 2005SGR-00889 of the Generalitat de Catalunya and
URJC-CM-2006-CET-0643  of the Universidad Rey Juan Carlos and Comunidad de Madrid is also acknowledged.

\section*{References}
\bibliography{synchro}

\end{document}